\documentclass[prd,aps,showpacs,nofootinbib,preprint,eqsecnum]{revtex4}

\usepackage{amsmath}
\usepackage{amssymb}
\usepackage{amsfonts}
\usepackage{graphicx,bm}
\usepackage{dcolumn}

\usepackage{color,amsxtra}
\usepackage{epsf}

\usepackage{enumerate}
\usepackage{hhline}
\usepackage{array}
\usepackage{tabularx}

\usepackage{subfigure}

\usepackage{fancyhdr}
\usepackage{mathrsfs}

\usepackage{amsmath}
\usepackage{amsthm}
\usepackage{amssymb}
\usepackage{amscd}
\usepackage[british]{babel}
\usepackage{babel}
\usepackage{graphicx}
\usepackage{psfrag}
\usepackage{epsfig}
\usepackage{rotating}
\usepackage{times}

\theoremstyle{plain}

\newtheorem{remark}{Remark}[section]

\newcommand{\boxend}{\flushright{$\Box$}}

\renewcommand{\tilde}{\widetilde}

\begin{document}

\title{How can holonomy corrections be introduced in  $f(R)$ gravity?}
\author{
Jaume de Haro$^{1, }$\footnote{E-mail address: jaime.haro@upc.edu},
}

\affiliation{
$^1$Departament de Matem\`atica Aplicada I, Universitat
Polit\`ecnica de Catalunya, Diagonal 647, 08028 Barcelona, Spain \\
}

\begin{abstract}
We  study the introduction of holonomy corrections in $f(R)$ gravity. We will show that there are infinitely many ways, as many as canonical transformations, to introduce
this kind of corrections, depending on the
canonical variables (two coordinates and its conjugate momenta) used to obtain the Hamiltonian. In each case, these  corrections lead, at effective level, to different
modified holonomy corrected Friedmann equations in $f(R)$ gravity, which are in practice analytically unworkable, i.e. only numerical analysis can be used to understand its dynamics.
Finally, we give arguments in favour of one preferred set of variables, the one that conformally maps $f(R)$ to Einstein gravity, because for these variables the dynamics of
the system has a clear physical meaning: the same as in standard Loop Quantum Cosmology, where the effective dynamics of a system can be analytically studied.
\end{abstract}


\maketitle

\section{Introduction}

It is well-known that, in general, $f(R)$ gravity 
(see \cite{no11} for a general introduction to $f(R)$ theories)
contains singularities, for example: all solutions of $R^2$ gravity in flat Friedmann-Lema{\^\i}tre-Robertson-Walker (FLRW)
are singular at early times (see for instance \cite{aho14}). A way to avoid these singularities is to introduce  holonomy corrections as in Loop Quantum Cosmology, because
this kind of corrections  provides a big bounce that avoids singularities like the Big Bang and Big Rip (see for example \cite{singh}).
Moreover, due to the big bounce, holonomy corrected $f(R)$ belongs into the set of  bouncing scenarios that could be a viable alternative to the inflationary paradigm
(see \cite{brandenberger} for a review of bouncing cosmologies).

The first attempt to
introduce  holonomy corrections in $f(R)$ gravity has been recently performed in \cite{zm11,zm11a}. The basic idea in those works is the conformal equivalence between $f(R)$ and
Einstein gravity, more precisely, the equivalence between
the dynamical equations of $f(R)$ gravity in the vacuum, obtained using as variables the coefficients of the metric, and the dynamical equations  in general relativity
when matter is represented by an scalar field, provided that the coefficients of a conformal metric were used as dynamical variables.

In cosmology, where the universe is approximately described by a flat FLRW geometry,
the introduction of holonomies in $f(R)$ gravity is greatly simplified because the only relevant geometrical variables are
the Hubble parameter and its first and second derivatives.
In this context, we study the different ways to extend LQC to $f(R)$ theories, i.e., the ways to introduce holonomy corrections in $f(R)$ gravity, and, as a consequence, the deduction of the
effective equations provided by these extensions.

These extensions will depend on the variables (two coordinates and two momenta) used to obtain the Hamiltonian of the system, and since there are infinitely many canonical transformations,
this means that there are infinite many ways to introduce holonomies in $f(R)$ gravity, leading to different effective holonomy corrected Friedmann equations in $f(R)$ gravity.
These modified Friedmann equations will depend, in a very complicated way, on the Hubble parameter $H$ and its first and second derivatives, meaning that the natural
phase space containing the orbits of the system is the plane $(H,\dot{H})$. These equations are more
complicated than the classical Friedmann equation for $f(R)$ gravity, which is already analytically unworkable, meaning that when one deals with holonomy corrections, in practice,
only numerical methods can be used to understand the dynamics of the system. However, form our point of view, the variables that conformally map $f(R)$ with Einstein gravity
are preferred in the sense that they provide a clear physical interpretation of the dynamical equations: the same as in Loop Quantum Cosmology, which allows us to understand the
dynamics in these coordinates and, with the help of this conformal map, to have a clear representation of the phase portrait in the plane $(H,\dot{H})$.

\section{$f(R)$ gravity ''a la Ostrogradsky''}
When one considers the flat FLRW geometry, the Lagrangian of
 $f(R)$ gravity in the vacuum is given by
\begin{eqnarray}\label{A1}
 {\mathcal L}(V,\dot{V},\ddot{V})=\frac{1}{2}Vf(R),
\end{eqnarray}
where $V=a^3$ is the volume and $R=\frac{2\ddot{V}}{V}-\frac{2\dot{V}^2}{3V^2}$ is the scalar curvature in terms of the volume and its derivatives.

Ostrogradsky's idea to obtain the Hamiltonian from a Lagrangian containing higher order derivatives is to introduce a Lagrange multiplier, namely $\mu$,
in the Lagrangian  as follows \cite{dsy}:
\begin{eqnarray}\label{A2}
 {\mathcal L}_1(V,\dot{V},\ddot{V},R)=\frac{1}{2}Vf(R)+\mu \left(\frac{2\ddot{V}}{V}-\frac{2\dot{V}^2}{3V^2}-R\right).
\end{eqnarray}

Extremization with respect to $R$ gives $\mu=\frac{1}{2}Vf_R(R)$, where we have introduced the notation $f_R(R)\equiv \partial_{R}f(R)$. In order to
remove the second derivative of $V$, we substract to the Lagrangian  the following
total derivative $\frac{d}{dt}\left(f_R(R) \dot{V}\right)$, which does not change the dynamics of the system,
and we replace the Lagrange multiplier $\mu$ by its value $\frac{1}{2}Vf_R(R)$, obtaining
\begin{eqnarray}\label{A3}
 \tilde{\mathcal L}(V,\dot{V},R,\dot{R})=\frac{1}{2}Vf(R)- \frac{1}{2}Vf_R(R)\left(\frac{2\dot{V}^2}{3V^2}+R\right)-f_{RR}(R)\dot{R} \dot{V}.
\end{eqnarray}

We can see that the Lagrangian  (\ref{A3}) depends on the variables $(V,R)$ and its first derivatives. The corresponding canonically conjugate momenta are
\begin{eqnarray}\label{A4}
 p_V\equiv \frac{\partial\tilde{\mathcal L}(V,\dot{V},R,\dot{R})}{\partial \dot{V}}=-\frac{2\dot{V}}{3V}f_R-f_{RR}\dot{R},\quad
 p_R\equiv \frac{\partial\tilde{\mathcal L}(V,\dot{V},R,\dot{R})}{\partial \dot{R}}=-f_{RR}\dot{V},
\end{eqnarray}
and the classical Hamiltonian  of the system becomes
\begin{eqnarray}\label{A5}
 {\mathcal H}_{c}(V,R,p_V,p_R)\equiv \dot{V}p_V+\dot{R}p_R-\tilde{\mathcal L}(V,\dot{V},R,\dot{R})
 =-\frac{p_Vp_R}{f_{RR}}+\frac{1}{3V}\frac{f_R}{f_{RR}^2}p_R^2+\frac{V}{2}\left(Rf_R-f\right).
\end{eqnarray}

The Hamiltonian constrain ${\mathcal H}_{c}(V,R,p_V,p_R)=0$ leads to the well-known modified Friedmann equation in $f(R)$ gravity
\begin{eqnarray}\label{A6}
 6f_{RR}\dot{R}H+6H^2f_R-(Rf_R-f)=0.
\end{eqnarray}

 Note that in Einstein gravity $f(R)=R$, one has $\frac{p_R}{f_{RR}}=-\dot{V}$ and $p_V=-\frac{2\dot{V}}{3V}=-2H$ and, thus, the classical Hamiltonian reduces to
 \begin{eqnarray}\label{A7}
  {\mathcal H}_{c}(V,p_V)=-\frac{3}{4}p_V^2V=-3H^2V.
 \end{eqnarray}

 \subsection{Canonical tranformations}
It's well known that $f(R)$ gravity is conformally equivalent to Einstein gravity (see for instance \cite{dsy}), which means that there exists a canonical transformation
${\mathcal T}$ of the form
\begin{eqnarray}\label{B1}
 {\mathcal T}:(V,R,p_V,p_R)\longmapsto (\bar{V},\bar{\phi},p_{\bar{V}},p_{\bar{\phi}}),
\end{eqnarray}
defined by
 \begin{eqnarray}\label{B2}
 {\mathcal T}(V,R,p_V,p_R)=\left(f_R^{3/2}{V},\sqrt{\frac{3}{2}}\ln{f_R},\frac{p_{{V}}}{f_R^{3/2}},\sqrt{\frac{3}{2}}\left(\frac{2}{3}\frac{p_Rf_R}{Vf_{RR}}-p_V \right)\right).
\end{eqnarray}

\begin{remark}
 The canonical transformation ${\mathcal T}$ can be obtained as follows: The Lagrangian (\ref{A1}) is equivalent to
 \begin{eqnarray}
  \bar{\mathcal L}=-\frac{({\bar V}')^2}{3\bar V}+\left(\frac{(\bar{\phi}')^2}{2}+W(\bar{\phi})\right)\bar{V},
 \end{eqnarray}
with  $W(\bar{\phi})=\frac{Rf_R-f}{2f_R^2}$, and $'\equiv \frac{d}{d\bar t}$ where we have introduced the conformal time $d\bar{t}= e^{\frac{\bar\phi}{\sqrt{6}}} dt$.
 And, finally, from this Lagrangian one obtains the conjugated momenta
 \begin{eqnarray}
  p_{\bar V}\equiv \frac{\partial \bar{\mathcal L}}{\partial\bar{V}'}, \quad p_{\bar \phi}\equiv \frac{\partial \bar{\mathcal L}}{\partial\bar{\phi}'}.
 \end{eqnarray}
\end{remark}

One can easily check that this transformation is canonical introducing the Poisson bracket
\begin{eqnarray}\label{B3}
 \left\{M,N\right\}\equiv \left(\partial_{p_V}M \partial_{V}N-\partial_{V}M \partial_{p_V}N\right)+\left(\partial_{p_R}M \partial_{R}N-\partial_{R}M \partial_{p_M}N\right),
\end{eqnarray}
a simple calculation yields 
\begin{eqnarray}\label{B4}
 \left\{\bar{V},p_{\bar{V}}\right\}=\left\{\bar{\phi},p_{\bar{\phi}}\right\}=-1,\quad \left\{\bar{V},\bar{\phi}\right\}=\left\{\bar{V},p_{\bar{\phi}}\right\}=
 \left\{p_{\bar{V}},\bar{\phi}\right\}=\left\{p_{\bar{V}},p_{\bar{\phi}}\right\}=
 0.
\end{eqnarray}

After Legendre's transformation the
  corresponding Hamiltonian in new variables is given by
 \begin{eqnarray}\label{B5}
  \bar{\mathcal H}_{c}(\bar{V},\bar{\phi},p_{\bar{V}},p_{\bar{\phi}})=\left(-\frac{3}{4}p_{\bar V}^2\bar{V}+\frac{p_{\bar{\phi}}^2}{2\bar{V}}+W(\bar{\phi})\bar{V}\right)
  e^{\frac{\bar\phi}{\sqrt{6}}},
 \end{eqnarray}
where $W(\bar{\phi})=\frac{Rf_R-f}{2f_R^2}$.

The Hamilton equations are $\dot{\mathcal A}=\left\{\bar{\mathcal H}_{c},{\mathcal A}\right\}$, where ${\mathcal A}=\bar{V},
\bar{\phi},p_{\bar{V}}$ and $p_{\bar{\phi}}$. Using the conformal time $\bar{t}$ introduced above, the
Hamilton equations become $\frac{d{\mathcal A}}{d\bar{t}}\equiv{\mathcal A}'=\left\{\tilde{\mathcal H}_{c},{\mathcal A}\right\}$, where the Hamiltonian
$\tilde{\mathcal H}_{c}$ has the more familiar form
\begin{eqnarray}\label{B6}
  \tilde{\mathcal H}_{c}(\bar{V},\bar{\phi},p_{\bar{V}},p_{\bar{\phi}})=-\frac{3}{4}p_{\bar V}^2\bar{V}+\frac{p_{\bar{\phi}}^2}{2\bar{V}}+W(\bar{\phi})\bar{V}.
 \end{eqnarray}

In fact, this Hamiltonian corresponds to a dynamical system given by an scalar field $\bar{\phi}$ under the action of the
potential $W(\bar{\phi})$ in Einstein Cosmology (EC). Introducing the energy density
\begin{eqnarray}\label{B7}
\bar{\rho}=\frac{p_{\bar{\phi}}^2}{2\bar{V}^2}+W(\bar{\phi}),
\end{eqnarray}
the Hamiltonian can be written as
\begin{eqnarray}\label{B8}
  \tilde{\mathcal H}_{c}(\bar{V},\bar{\phi},p_{\bar{V}},p_{\bar{\phi}})=-\frac{3}{4}p_{\bar V}^2\bar{V}+\bar{\rho}\bar{V}.
 \end{eqnarray}

Hamiltonian (\ref{B8}) shows the canonical equivalence between $f(R)$ and Einstein gravity. Moreover, due to the conformal change
of variable $\bar{V}=f_R^{3/2}{V}$ (the first component on the right hand side of (\ref{B2})), one can argue that
$f(R)$ and Einstein gravity are conformally equivalent.

Note also that when $f(R)=R$,  $p_{\bar{V}}$ reduces to $-2H$, and  Hamiltonian (\ref{B8}) reduces to (\ref{A7}).

To end this Section,
a final remark is in order: Since there are infinitely many canonical transformations, this means that $f(R)$ gravity could be formulated using infinitely many sets of variables (two coordinates
and their corresponding momenta). Note that  some of these sets of variables will be meaningless physically speaking, because they are built using a combination of
both coordinates and momenta, giving new quantities with a very difficult physical interpretation.
 Moreover, we will show that the introduction of holonomy effects will depend on the set of variables used, i.e., there will be
infinitely many ways to introduce
holonomy effects in $f(R)$ gravity, and consequently, there will be infinitely many different effective holonomy corrected Friedmann equations in $f(R)$ gravity.

 \section{Introduction of holonomy corrections}
 For the flat FLRW geometry, when one deals with Einstein gravity to  introduce holonomies:  first of all one can consider the variable $\beta=-\frac{\gamma}{2}p_V=\gamma H$
 (\cite{s09})
 where $\gamma$ is the Barbero-Immirzi parameter. In term of $\beta$
 the Hamiltonian (\ref{A7})
becomes ${\mathcal H}_{c}(V,\beta)=-\frac{3\beta^2}{\gamma^2}V$.
However,
in Loop Quantum Cosmology, due to the discrete nature of space,  the quantum operator $\hat{\beta}$ is not well defined
(see for instance \cite{h12} or \cite{as11} for a status report). Then, in order to build the quantum theory, one needs to
re-define the gravitational part of the Hamiltonian. To be precise,  we will consider the holonomies $
h_j(\lambda)\equiv e^{-i\frac{\lambda \beta}{2}\sigma_j}$, where
$\sigma_j$ are the Pauli matrices and $\lambda$ is the square root of the minimum eigenvalue of the area operator in Loop Quantum Gravity.
Since $\beta^2$ does not have a well-defined quantum operator, to construct a consistent  quantum Hamiltonian operator, one needs
an almost periodic function that approaches $\beta^2$ for small values of $\beta$.
This can be done using
the general formulae of loop quantum gravity to
obtain the holonomy corrected Hamiltonian
\begin{eqnarray}\label{ham}
&& \hspace*{-5mm}
{\mathcal H}_{hc}(V,\beta)\equiv-\frac{2 {V}}{\gamma^3 \lambda^3}
\sum_{i,j,k}\varepsilon^{ijk} Tr\left[
h_i(\lambda)h_j(\lambda)h_i^{-1}(\lambda)
 h_j^{-1}(\lambda)h_k(\lambda)\{h_k^{-1}(\lambda),{V}\}\right],
\end{eqnarray}
which
 captures the underlying loop quantum dynamics (see for instance \cite{abl03, aps06}).

A simple calculation proves  \cite{bo08,he10,dmw09} that (\ref{ham}) aquires the simple form
\begin{eqnarray}\label{A8}
  {\mathcal H}_{hc}(V,\beta)=-3\frac{\sin^2(\lambda \beta)}{\lambda^2\gamma^2}V,
 \end{eqnarray}
which shows that, at effective level,
 holonomy effects can be introduced performing  the replacement
$\beta\rightarrow \frac{\sin(\lambda \beta)}{\lambda}$ (equivalently, $p_V\rightarrow -\frac{2\sin(\lambda \beta)}{\lambda\gamma}$).

To obtain the  holonomy corrected Friedmann equation one has  to use the full Hamiltonian
\begin{eqnarray}\label{A9}{\mathcal H}_{full}(V,\beta) =-3\frac{\sin^2(\lambda \beta)}{\lambda^2\gamma^2}V+\rho V,\end{eqnarray}
to calculate the
Hamilton
equation
\begin{eqnarray}\label{A10}
 \dot{V}=-\frac{\gamma}{2}\frac{\partial{\mathcal H}_{full}(V,\beta) }{\partial \beta}=-3V\frac{\sin(2\lambda \beta)}{2\lambda\gamma},
\end{eqnarray}
that, together with the Hamiltonian constrain ${\mathcal H}_{full}(V,\beta) =0$, leads to the well-known  modified Friedmann equation
\begin{eqnarray}\label{A11}
 H^2=\frac{\rho}{3}\left(1-\frac{\rho}{\rho_c}\right),
\end{eqnarray}
where $\rho_c\equiv\frac{3}{\lambda^2\gamma^2}$ is the so-called {\it critical density}.

\subsection{Holonomy corrected  $f(R)$ gravity}
To introduce the holonomy correction in  general $f(R)$
we will adopt the following recipe:
In  analogy with the linear case $f(R)=R$, we will replace the momentum that in the linear
case corresponds to $-2\frac{\beta}{\gamma}$ by $-\frac{2\sin(\lambda \beta)}{\lambda\gamma}$. For example,
if the variables $(V,R,p_V,p_R)$ are used, we will perform
in Hamiltonian
(\ref{A5}) the replacement  $p_V\rightarrow -\frac{2\sin(\lambda \beta)}{\lambda\gamma}$, and
if we consider the variables $(\bar{V},\bar{\phi},p_{\bar{V}},p_{\bar{\phi}})$, we will replace
in  the Hamiltonian  (\ref{B8})
$p_{\bar V}$ by $-\frac{2\sin(\lambda \beta)}{\lambda\gamma}$, because, in both cases, when
 $f(R)=R$ one has $p_V=p_{\bar{V}}=-2\frac{\beta}{\gamma}$.

 It is important to realize that
 this way to introduce holonomy corrections will depend on the set of variables used to formulate  $f(R)$. We can prove it, introducing holonomy corrections in Hamiltonians
 (\ref{A5}) and (\ref{B8}), and showing that these corrections lead to different differential equations.

\begin{enumerate}\item
First of all we deal with  the Hamiltonian (\ref{A5}). After the replacement one gets
\begin{eqnarray}\label{A12}
 {\mathcal H}_{hc}
 =\frac{2\sin(\lambda\beta)p_R}{\lambda\gamma f_{RR}}+\frac{1}{3V}\frac{f_R}{f_{RR}^2}p_R^2+\frac{V}{2}\left(Rf_R-f\right).
\end{eqnarray}

The Hamilton equations
\begin{eqnarray}\label{A13}
\dot{V}=-\frac{\gamma}{2}\frac{\partial{\mathcal H}_{hc} }{\partial \beta}= -\frac{\cos(\lambda\beta)p_R}{f_{RR}},\quad
\dot{R}=\frac{\partial{\mathcal H}_{hc} }{\partial p_R}=\frac{2\sin(\lambda\beta)}{\lambda\gamma f_{RR}}+\frac{2}{3V}\frac{f_R}{f_{RR}^2}p_R,
\end{eqnarray}
together with the Hamiltonian constrain ${\mathcal H}_{hc}=0$, have to be used to obtain a relation of the form $F(H,R,\dot{R})=0$ which corresponds to the modified Friedmann
equation in $f(R)$ gravity containing holonomy corrections.

To do that, first of all we introduce the notation $p_R=V\tilde{p}_R$. Then, equations (\ref{A13}) become
\begin{eqnarray}\label{A14}
 \sin^2(\lambda\beta)=1- \frac{9H^2f_{RR}^2}{\tilde{p}_R^2},
\end{eqnarray}
and
\begin{eqnarray}\label{A15}
\sin^2(\lambda\beta)=\frac{3}{4\rho_c}\left( \dot{R}^2f_{RR}^2+\frac{2}{3}f_R(Rf_R-f)\right),
\end{eqnarray}
where to obtain this last equation we have used the Hamiltonian constrain. Equalising both equations one has
\begin{eqnarray}\label{A16}
 \tilde{p}_R^2=\frac{9H^2f_{RR}^2}{1-\frac{3A}{4\rho_c}},
\end{eqnarray}
where $A= f^2_{RR}\dot{R}^2+\frac{2}{3}f_R(Rf_R-f)$.

Finally, inserting (\ref{A14}) and (\ref{A16}) in the square of the Hamiltonian constrain one gets
\begin{eqnarray}\label{A17}
  9H^2f_{RR}^2\dot{R}^2\left(1-\frac{3A}{4\rho_c}\right)=\frac{1}{4}\left[6H^2f_R-(Rf_R-f)\left(1-\frac{3A}{4\rho_c}\right)\right]^2.
\end{eqnarray}

\begin{remark}
 It is important to realize that, when $f(R)=R$, equation (\ref{A17})  leads to  equation (\ref{A11}). To prove that, one only has to introduce an energy
 density $\rho$, which can be done replacing $(Rf_R-f)$ by $(Rf_R-f-2\rho)$ in (\ref{A17}).
\end{remark}

\item Finally, we deal with the Hamiltonian (\ref{B8}). After the replacement $p_{\bar V}\rightarrow -\frac{2\sin(\lambda \beta)}{\lambda\gamma}$
in (\ref{B8}), one obtains
\begin{eqnarray}
  \tilde{\mathcal H}_{hc}=-3\frac{\sin^2(\lambda \beta)}{\lambda^2\gamma^2}\bar{V}+\bar{\rho}\bar{V}.
 \end{eqnarray}

Using the Hamilton equation $\bar{V}'=\left\{\tilde{\mathcal H}_{hc},\bar{V}\right\}$ and the Hamiltonian constrain
$\tilde{\mathcal H}_{hc}=0$ one obtains the following holonomy  corrected Friedmann equation
\begin{eqnarray}\label{Friedmann}
 \bar{H}^2=\frac{\bar\rho}{3}\left(1-\frac{\bar\rho}{\rho_c}\right),
\end{eqnarray}
where $\bar{H}=\frac{1}{3}\frac{\bar{V}'}{\bar{V}}$ is a conformal Hubble parameter, which is exactly the same equation as (\ref{A11}) but with the new variables.

Using the variables $(V,\dot{V},R,\dot{R})$ this equation becomes (note that ${H}=\frac{1}{3}\frac{\dot{V}}{{V}}$)
\begin{eqnarray}\label{A18}
  6f_{RR}\dot{R}H+6H^2f_R -(Rf_R-f)=
-\frac{9A^2}{8f_R^4{\rho}_c},
\end{eqnarray}
\end{enumerate}
which is completely different from (\ref{A17}), because it contains a linear term on $\dot{R}$. In the same way, using other different variables we will obtain different effective
holonomy corrected Friedmann equations in $f(R)$ gravity.

What is important in all of these modified Friedmann equations obtained using different variables, is that all of them are autonomous second order differential equations in $H$, that is,
they contain $H, \dot{H}$ and $\ddot{H}$,
and do not contain explicitly the time.
This means that the dynamical
system is contained in the phase space $(H,\dot{H})$. The problem with this kind of equations is that, in general, they are analytically unworkable and only numerical computations can be
performed.

However,  equation (\ref{A18}) has an advantage with respect to all the other formulations of holonomy corrected $f(R)$ gravity: The relation
\begin{eqnarray}\label{rel1}
H=e^{\frac{\bar\phi}{\sqrt{6}}}\left(\bar{H}-\frac{\bar{\phi}'}{\sqrt{6}}\right),\quad
\dot{H}=\frac{\bar{\phi}'e^{\sqrt{\frac{2}{3}}\bar\phi}}{\sqrt{6}}\left(\bar{H}-\frac{\bar{\phi}'}{\sqrt{6}}\right)+
e^{\sqrt{\frac{2}{3}}\bar\phi}\left(\bar{H}'-
\frac{\bar{\phi}''}{\sqrt{6}}\right),
\end{eqnarray}
 provides a map between the planes $(H,\dot{H})$ and $(\bar{\phi},\bar{\phi}')$, where, as we will show,
 the dynamics of the system is easier to understand. Effectively, the energy density $\bar{p}$ has the following form in terms of
$(\bar{\phi},\bar{\phi}')$
\begin{eqnarray}
 \bar{\rho}=\frac{(\bar{\phi}')^2}{2}+W(\bar{\phi}).
\end{eqnarray}

Then, the holonomy corrected Friedmann equation (\ref{Friedmann}) relates $\bar{H}$ with $(\bar{\phi},\bar{\phi}')$, and
the dependence of $\bar{H}'$ and $\bar{\phi}''$  with respect to $(\bar{\phi},\bar{\phi}')$  is obtained through the Raychaudhuri
\begin{eqnarray}\label{Raychaudhuri}
 \bar{H}'=-\frac{1}{2}\left(1-\frac{2\bar{\rho}}{\rho_c} \right)(\bar{\phi}')^2,
\end{eqnarray}
and  conservation equations
\begin{eqnarray}\label{conservation}
 \bar{\rho}'=-3\bar{H}(\bar{\phi}')^2 \Longleftrightarrow \bar{\phi}''+3\bar{H}\bar{\phi}'+\partial_{\bar{\phi}}W(\bar{\phi})=0.
\end{eqnarray}

From these equations, the second term in (\ref{rel1})  becomes
\begin{eqnarray}\label{rel1a}\hspace{-0.75cm}
\dot{H}=\frac{\bar{\phi}'e^{\sqrt{\frac{2}{3}}\bar\phi}}{\sqrt{6}}\left(\bar{H}-\frac{\bar{\phi}'}{\sqrt{6}}\right)+
e^{\sqrt{\frac{2}{3}}\bar\phi}\left(
-\frac{1}{2}\left(1-\frac{(\bar{\phi}')^2+2 W(\bar{\phi})}{\rho_c}\right)(\bar{\phi}')^2
+
\frac{3\bar{H}\bar{\phi}'+\partial_{\bar{\phi}}W(\bar{\phi})}{\sqrt{6}}\right),
\end{eqnarray}
which only depends on the variables $(\bar{\phi},\bar{\phi}')$.

Consequently, the dynamics in phase space $(H,\dot{H})$ is obtained
from the one in the plane  $(\bar{\phi},\bar{\phi}')$. Working with these variables equation (\ref{Friedmann}) shows that
the universe moves along an ellipse in the plane $(\bar{H},\bar{\rho})$, like in standard Loop Quantum Cosmology. Moreover the conservation equation (\ref{conservation})
shows that the movement is clockwise from the contracting ($\bar{H}<0$) to the expanding ($\bar{H}>0$) phase in these variables,
 bouncing at $\bar{\rho}=\rho_c$.

 The dynamics in the phase space $(\bar{\phi},\bar{\phi}')$  is obtained from the conservation equation (\ref{conservation}) which in the contracting
 phase ($\bar{H}<0$) is given by
 \begin{eqnarray}\label{KG1}
 \bar{\phi}''-3\sqrt{\frac{\frac{(\bar{\phi}')^2}{2}+W(\bar{\phi})}{3}
 \left(1- \frac{\frac{(\bar{\phi}')^2}{2}+W(\bar{\phi})}{\rho_c} \right)
 }\bar{\phi}'+\partial_{\bar{\phi}}W(\bar{\phi})=0,
 \end{eqnarray}
and in the expanding one ($\bar{H}>0$)  by
 \begin{eqnarray}\label{KG2}
 \bar{\phi}''+3\sqrt{\frac{\frac{(\bar{\phi}')^2}{2}+W(\bar{\phi})}{3}
 \left(1- \frac{\frac{(\bar{\phi}')^2}{2}+W(\bar{\phi})}{\rho_c} \right)
 }\bar{\phi}'+\partial_{\bar{\phi}}W(\bar{\phi})=0.
 \end{eqnarray}

Strictly speaking equations (\ref{KG1}) and (\ref{KG2}) depict two autonomous dynamical systems in the phase space $(\bar{\phi},\bar{\phi}')$, meaning that orbits in the contracting phase ($\bar{H}<0$) 
intersect with the ones in the expanding phase ($\bar{H}>0$). The numerical
integration of this system, and thus, the numerical phase portrait could be done as follows:  Given an initial condition $(\bar{\phi}_0,\bar{\phi}'_0)$, one integrates forward in time equation (\ref{KG1})
for this initial condition. Then, the orbit could  move to a critical point of the system
$(\bar{\phi}_c,0)$ where $\bar{\phi}_c$ is a solution of the equation $\partial_{\bar{\phi}}W(\bar{\phi})=0$, or  it hits tangentially
the curve $\bar{\rho}=\rho_c$ at some point  $(\bar{\phi}_1,\bar{\phi}'_1)$. In the latter case one has to continue the orbit
integrating  forward in time (\ref{KG2}) for this new initial condition. In this way one obtains the phase portrait
in the plane $(\bar{\phi},\bar{\phi}')$, and from the map (\ref{rel1}) one finally obtains the phase portrait in the plane
 $(H,\dot{H})$. In fact,
a realistic application of
this method has recently been performed in \cite{aho14} in order to obtain the phase portrait  for
 $R+\alpha R^2$ Loop Quantum Cosmology.

 \vspace{0.5cm}
{\bf Acknowledgments:}
The author  thanks Professors
S.D. Odintsov and J. Amor\'os for helpful discussions.
This investigation has been supported in part by MINECO (Spain)
MTM2011-27739-C04-01, and by AGAUR (Generalitat de
Catalunya), Contract No. 2009SGR-345.

\end{document}